# Growth mechanisms and structure of fullerene-like carbon-based thin films: superelastic materials for tribological applications

R. Gago,[1] G. Abrasonis,[2] I. Jiménez,[3] and W. Möller[2]

[1] Centro de Micro-Análisis de Materiales, Universidad Autónoma de Madrid, E-28049 Madrid, Spain.
[2] Institute of Ion Beam Physics and Materials Research, Forschungszentrum Dresden-Rossendorf, PF510119, D-01314 Dresden, Germany.
[3] Instituto de Ciencia de Materiales de Madrid, Consejo Superior de Investigaciones Científicas, E-28049 Madrid, Spain.

**Abstract:**

In this chapter we review our findings on the bonding structure and growth mechanisms of carbon-based thin solid films with fullerene-like (FL) microstructure. The so-called FL arrangements arise from the curvature and cross-linking of basal planes in graphitic-like structures, partially resembling that of molecular fullerenes. This three-dimensional superstructure takes advantage of the strength of planar π bonds in sp$^2$ hybrids and confers the material interesting mechanical properties, such as high hardness, high elastic recovery, low-friction and wear-resistance. These properties can be tailored by controlling the curvature, size and connectivity of the FL arrangements, making these materials promising coatings for tribological applications. We have focused our interest mostly on carbon nitride ($CN_x$) since nitrogen promotes the formation of FL arrangements at low substrate temperatures and they are emerging over pure carbon coatings in tribological applications such as protective overcoats in magnetic hard disks. We address structural issues such as origin of plane curvature, nature of the cross-linking sites and sp$^2$ clustering, together with growth mechanisms based on the role of film-forming precursors, chemical re-sputtering or concurrent ion assistance during growth.

## 1. INTRODUCTION

During the last years there has been a growing interest on carbon-based nanostructures like fullerenes, nanotubes and nano-onions due to their remarkable and tunable mechanical, electronic and electrochemical properties [1]. The common characteristic of these nanostructures is the presence of graphitic atomic arrangements with in-plane curvature, giving rise to 3-dimensional (3D) structure of $sp^2$-coordinated atoms. These 3D units constitute closed cages in the case of spherical (hetero-)fullerene molecules and onion-like carbon.

All-$sp^2$ 3D curved structures have been also reported in graphitic carbon-based thin solid films of pure carbon (C) [2], carbon nitride ($CN_x$) [3] and hexagonal boron carbon nitride (BCN) [4]. These compounds have been named as fullerene-like (FL) materials due to the presence of curved and cross-linked basal planes, resembling the structure of fullerene molecules. The presence of FL arrangements within the solid is normally assessed by high resolution transmission electron microscopy (HRTEM). The typical micrograph representing such structure in $CN_x$ grown is depicted in Figure 1. As observed, the degree of in-plane curvature can be considerable inducing even the formation of onion-like shells, which are cross-linked to yield a solid-phase [5].

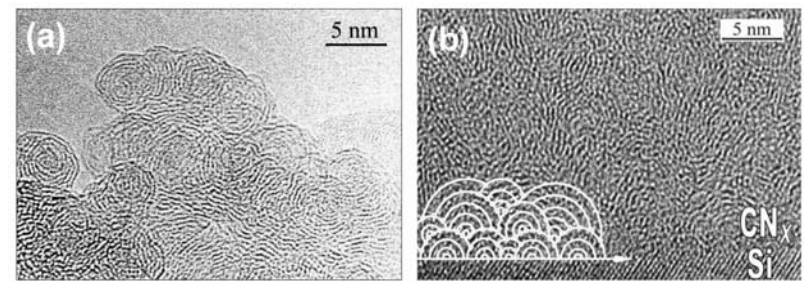

**Figure 1.** HRTEM plan-view (a) and cross-sectional (b) image of FL arrangements in a $CN_x$ thin film grown by DC-MS. The interconnectivity of the onion-like shells yields the formation of a solid phase (taken from Refs. 5 and 6 with permission of Zsolt Czigány, Jörg Neidhardt, and Lars Hultman).

The main interest on FL arrangements is triggered by the extraordinary mechanical properties that they confer to the material. The 3D FL structure takes advantage of the strong bond between $sp^2$ hybridized atoms. Note that the in-plane bond in graphite is actually stronger than in the corresponding $sp^3$ allotrope (diamond), whereas the overall strength of graphite is limited by the anisotropic character imposed with its laminar structure. The presence of curvature introduces a high elastic recovery (>80%) under mechanical deformation with the peculiarity that it is attained simultaneously with moderate hardness (15-20 GPa). The hardening mechanism is induced by the interconnectivity or cross-linking between the curved planes. Therefore, FL nanostructured coatings are compliant and tough, making them ideal candidates as materials for tribological applications [6]. Moreover, these mechanical properties are tunable since they can be tailored by the orientation, corrugation, folding, and cross-linking of the basal planes [7]. As an example of the last statement, the film properties and microstructure for FL-$CN_x$ grown by reactive ($N_2$/Ar) DC magnetron sputtering (DC-MS) as a function of deposition parameters can be seen in Figure 2.

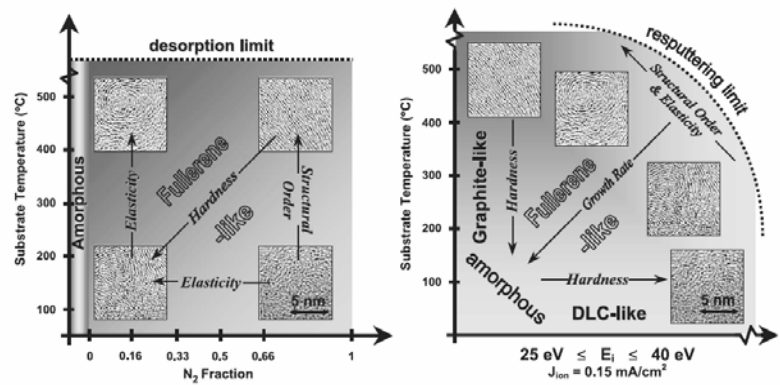

**Figure 2.** Zone diagram showing the microstructure and mechanical properties of FL-$CN_x$ grown by DC-MS as a function of substrate temperature, $N_2$ content in the Ar/$N_2$ gas mixture and ion assistance energy (given by the substrate polarization or bias voltage) (taken from Ref. 7, courtesy of Jörg Neidhardt and Lars Hultman).

In the case of carbon-based thin films, the formation of FL arrangement has been mainly focused on $CN_x$ materials. This is due to the fact that N promotes the formation of FL arrangements at lower substrate temperatures as compared to pure C films [8]. The predilection for $CN_x$ materials is also triggered by their superior tribological properties over C films, which are even enhanced upon promotion of FL arrangements [7]. An evidence of this trend is that amorphous $CN_x$ films are now replacing commercial diamond-like carbon (DLC) coatings as protective overcoats on magnetic storage media and read and write heads [9]. Finally, FL arrangements are not restricted to carbonaceous solids, but may be found in other laminar materials [10]. In this line, recent interest is being devoted to the synthesis of inorganic fullerene-like nanoparticles, like $WS_2$ or $MoS_2$ [11]. Recently, FL thin solid films of $MoS_2$ have been reported with extraordinary tribological properties such as ultra-low friction and wear [12].

## 2. OPEN ISSUES IN FULLERENE-LIKE THIN SOLID FILMS

### 2.1. Role of nitrogen incorporation in graphitic $CN_x$ solids

Despite the successful synthesis of C and $CN_x$ materials with FL structure, the atomic arrangements leading to such a microstructure are a matter of debate and still not fully understood. In particular, the mechanisms inducing curvature and the nature of the cross-linking sites are still unclear. As already mentioned, in the case of pure C films, the window of processing parameters yielding a FL structure is very narrow, with the FL structure being difficult to control. However, in $CN_x$ the formation and control of the FL structure is much easier since, for example, occurs at lower substrate temperatures. This suggests that N incorporation plays a key role for the promotion of buckling of the graphene sheets [13].

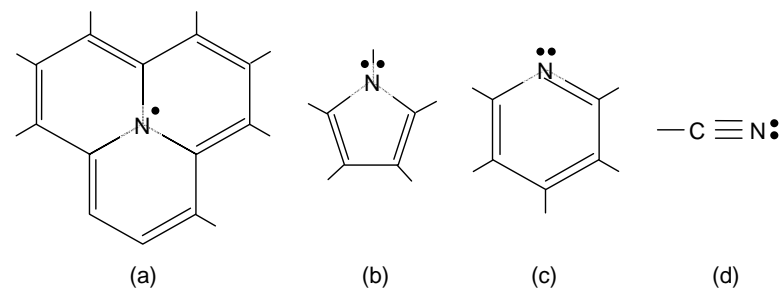

**Figure 3.** Different N bonding environments in a graphitic network with decreasing coordination: (a) substitutional (3-fold), (b) pyrrole-like (3-fold), (c) pyridine-like (2-fold) and (c) nitrile-like or cyanogen-like (1-fold). Solid dots indicate the number of non-binding electrons in the valence shell whereas double lines indicate conjugation of electrons inside the rings.

The incorporation of N in a graphitic C matrix induces significant changes in the film microstructure since it affects the local bonding structure and the $sp^2$ clustering process.

The N-induced modification can be attributed to changes in the bonding length and bond angle distortions due to the formation of C-N bonds instead of C-C and, additionally, to the capability of N to accommodate different local atomic arrangements and coordination such as substitutional sites in graphite, pyrrole-like, pyridine-like or nitrile-like configurations [14,15]. These different bonding environments can be identified in Figure 3.

In analogy with fullerene molecules, the bending of the basal planes has been initially attributed to formation of odd-member ring defects (i.e. pentagons or heptagons) in the graphitic structure, i.e. pyrrole-like environments (see Fig. 3(b)). These assumptions are supported by theoretical calculations by Stafström [16], showing that N incorporation can reduce the energy barrier required to create pentagon defects. However, in general, FL materials present a curvature of the basal planes smaller (larger radius of curvature) than in $C_{60}$ molecules, implying that the origin of bent planes can differ from the presence of odd-member rings. In this sense, another plausible folding mechanism of the graphene sheets considers corrugation induced by non-planar N atoms in graphitic substitutional sites (see Fig. 3(a)) [17]. These two folding pathways are schematically illustrated in the evolutions depicted in Figure 4.

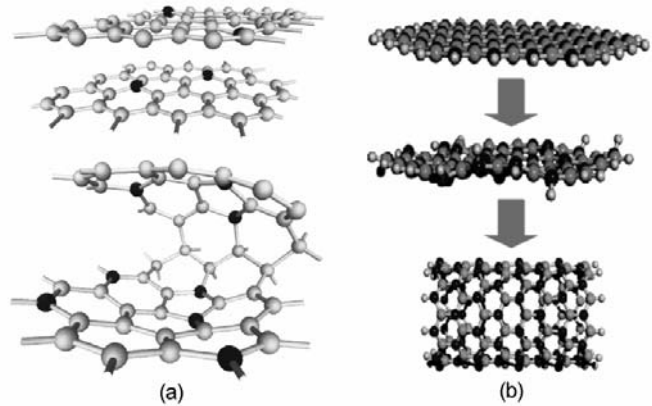

**Figure 4.** Models for N-induced curvature due to (a) formation of pentagons (adapted from Ref. 13) and (b) corrugation by N in substitutional graphite sites which evolves towards a 3D molecular structure (adapted from Ref. 17). In both structures, N atoms are displayed in a darker color.

However, the sole N incorporation can not explain the formation and evolution of FL arrangements since there is not a direct correlation between microstructure and the N content in the films [22]. That is, $CN_x$ with identical composition can appear as a hexagonal layered structure with almost flat basal planes, or with the highly developed aforementioned FL structure. From the bonding standpoint, there are no significant differences between hexagonal and FL structures, as indicated by spectroscopic results yielding similar signals for both types of $CN_x$ [18,19,20,21]. Hence, FL-$CN_x$ is not originated from a specific bonding environment that is absent in flat structures, but appears due to a subtle balance between various coordination possibilities. Therefore, the structure evolution has to be interpreted as a promotion of favourable bonding environments during the growth process. For example, the dominance of three-fold N in substitutional or pyrrole-like sites should be relevant for the development of FL microstructure as responsible for buckling in the grapheme sheets [3,17]. On the other hand, pyridine- and nitrile-like environments are terminating bonding configurations that may hinder the clustering evolution [18,19]. However, the presence of these terminating sites may be also necessary to accommodate the intrinsic stress arose by FL arrangements [20,21].

The theoretical results from Stafström [16] also indicate that N incorporation favors the formation of reactive C-$sp^3$ in adjacent sites. This prediction has been used to support the hypothesis that C-$sp^3$ atoms act as cross-linking sites [3]. However, contrary to this model, the bonding structure of DC-MS $CN_x$ films is experimentally found to be composed of ~100% $sp^2$ hybrids [19,20], without any evidence of significant $sp^3$-C formation. This controversy leaves this issue unclear and points out the eventual C $sp^2$-based [22] or N-based [23] cross-linkage.

### 2.2. Growth mechanisms of FL-$CN_x$

Most of the reports regarding the growth of FL materials in thin film form have been focused on FL-$CN_x$ synthesized by reactive DC-MS [6]. Apart from the role of N for promoting FL environments, this limitation is mainly due to the fact that, so far, this is the only technique able to produce well-structured FL-$CN_x$ thin films. Recently, pulsed laser deposition (PLD) resulted also in partially structured films [24]. Regarding the conditions for FL-$CN_x$ nucleation and subsequent growth, low-energy (<100 eV) ion bombardment and moderate substrate temperatures (400-800 K) have been shown to be essential requisites [13,25]. These growth conditions may be necessary to increase the surface mobility and reactivity but also to enable selective etching of less-favorable bonding environments without detrimental ion-induced damage (amorphization) for the evolution of the microstructure.

In the growth of $CN_x$ films (not limited to the case of FL-$CN_x$), a relevant factor is the presence of "chemical sputtering" [26,27,28]. This process implies the formation of volatile species at the substrate surface that results in a lower effective growth rate than the corresponding to the flux of incoming precursor species. This process is thermally activated and, therefore, it is promoted at high growth temperature [29]. Under such circumstance, no effective growth rate may be even attained. As a general trend, in the growth of $CN_x$ materials under energetic deposition techniques, the formation of CN volatile species at the substrate is related to the limitation in the N content, which rarely exceeds ~30 at. %. This chemical sputtering has also implications in the microstructure evolution of $CN_x$ solids [30].

The limited number of thin film growth techniques capable of FL-$CN_x$ synthesis suggests the there are special requirements related to the incorporation of C and N precursors. The analysis of the emitted species from the sputtering target during the growth of $CN_x$ films by MS indicates that, as occurs in the substrate surface, there is a non-negligible formation of $C_xN_y$ precursors due to the nitridation and chemical sputtering of the target [31]. This finding has been used to launch the hypothesis that pre-formed $C_xN_y$ moieties from the sputtering target may have eventual relevance as film-forming species. In this way, these precursors may imprint their structure on the evolving microstructure. First-principle calculations have shown that the incorporation of CN dimers and cyanogen molecules promote an initial stabilization of pentagon defects needed to induce FL arrangements [32].

In conclusion, the growth of FL-CNx has to be understood as an interplay between incoming hyperthermal atomic and polyatomic species, low-energy ion bombardment and desorption of volatile species due to the chemical sputtering. Here, physical sputtering does not to play a significant role due to the low energies (<100eV) required to preclude amorphization. The relative relevance of the mechanisms involved and their implications in such a complex growth process is still to be fully determined. Figure 5 shows schematically that the formation of FL-$CN_x$ occurs as the interplay between these mechanisms.

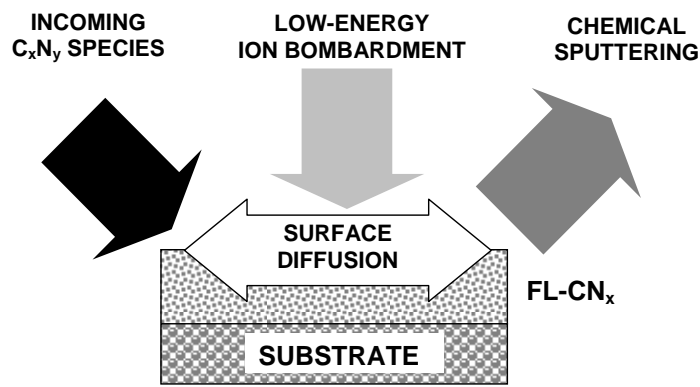

**Figure 5.** The complex growth of FL-$CN_x$ thin solid films as the interplay between incoming precursors (atomic and polyatomic $C_xN_y$ species), surface diffusion processes (temperature and/or ion induced), simultaneous low-energy ion bombardment and desorption (temperature activated) of volatile compounds (chemical sputtering).

## 3. MOTIVATION OF THIS CHAPTER

As shown in the previous section, there are unresolved issues regarding the atomic structure and growth mechanisms responsible for the formation of FL arrangements in graphitic networks. The posed questions are relevant to improve the control over the material properties and to understand the fundamental mechanisms leading to such kind of microstructure.

A critical point to study FL materials is that a direct observation of FL arrangements is limited to the use of HRTEM. However, the information that can be extracted from HRTEM is not straight-forward since preparation of TEM specimens may mask the observation of FL arrangements. For example, during ion beam milling for cross-section imaging preparation, high ion bombardment may destroy the structure due to surface amorphization. These drawbacks may be overcome by the reduction of the ion bombardment in the final steps of the thinning process (down to a few hundreds of eV instead of several keV) or by studying plan-view specimens when deposition on soluble substrates is possible [33]. Another problem in HRTEM analysis comes from projection artifacts, which may preclude the resolution of structural features. This can be partially solved by considering selective area electron diffraction (SAED), since any overlapping does not affect the characteristic lattice spacing in the diffraction pattern and information on the degree of ordering can be derived from the brightness and width of the diffraction pattern [34]. Finally, an additional drawback of TEM measurements for studying thin solid films is the destructive character entailed.

Further identification of FL arrangements with additional characterisation tools is desirable, especially if they are not destructive for the sample. This issue has been addressed by spectroscopic methods such as nuclear magnetic resonance (NMR) [19,20,35], x-ray photoemission spectroscopy (XPS) [34], electron energy loss spectroscopy [36] and x-ray absorption near-edge spectroscopy (XANES) [18,21,37] by correlation of microstructural features and the promotion of certain local bonding environments. Among them, NMR and XANES have been proved to be the most powerful ones in detecting and resolving the various bonding environments present in $CN_x$ solids. NMR has shown clearly that graphitic $CN_x$ contain multiple bonding environments upon N incorporation, including different types of pyridine-like coordination, nitriles, pyrroles and other organic functionalities [19,20]. The main drawback of NMR is that it requires detaching and powderizing (i.e., destructive character) the coating until reaching a relatively large amount of material (a few mg). XANES, instead, can be performed on a single coating without any sample preparation.

In this chapter, section 4 presents a detail XANES study of the bonding structure of FL-$CN_x$. In particular, we have studied $CN_x$ films grown by DC-MS with varying degree of FL character (in terms of plane alignment, extension and cross-linking) by tuning the proper growth parameters. The contribution of the different bonding environments has been correlated with the formation and promotion of FL arrangements. In this way, we can get further insight in the mechanisms for buckling of graphene-sheets or cross-linking sites upon N incorporation.

Regarding the growth mechanism, the relative contribution of the different bonding configurations and how they assemble will define the final film microstructure and, hence, the material properties. This issue depends strongly on the way that N is introduced into the C network, i.e. on the $CN_x$ synthesis process. Therefore, the comparison of the different bonding structures for films grown under different conditions can shed some light on the mechanisms leading to the different microstructures. In section 5 of this chapter, we present a comparative study with different $CN_x$ growth methods that partially mimic the conditions attained during MS, mainly in the frame of film precursors and ion assistance during growth. In this way, we aim at discerning the mechanisms operating during the evolution of FL arrangements. The identification of spectral signatures related with the formation and promotion of FL arrangements described in section 4 will be very useful for the conclusions of this study.

## 4. BONDING STRUCTURE OF FL-$CN_x$ BY XANES

### 4.1. Fundamentals of the XANES technique

X-ray absorption (XAS) features near the adsorption edge, normally referred as XANES or NEXAFS (near-edge x-ray absorption fine structure) spectroscopy, provide information about the K-shell structure with elemental sensitivity and, hence, is very sensitive to local bonding arrangements [38]. XANES detects the transitions from core-electrons to the conduction band of the solid, providing an image of the density of unoccupied states. Therefore, XANES provides a similar information to EELS or ELNES (energy loss near-edge spectroscopy), although at present the energy resolution of XAS studies is typically in the range of 10 to 100 meV, i.e. ten times better than in the electron energy loss experiments.

The XANES experiments discussed here were performed in the total electron yield (TEY) mode, consisting in the measurement of the photocurrent produced when the sample surface is illuminated with monochromatic x-ray light from a synchrotron ring. The spectrum is acquired by sweeping the photon energy in the desirable range and measuring the TEY signal. The XANES-TEY spectrum can be considered as the plot of the absorbance as a function of energy with good approximation. These experiments were performed in the SuperACO-LURE (Orsay, France), SSRL (Stanford, CA, USA) and BESSY-II (Berlin, Germany) synchrotron facilities.

## 4.2. XANES of carbonaceous materials

XANES is very suitable to study carbon-based materials due to the distinct features obtained for $sp^2$ and $sp^3$ hybrids. For this reason, it is one of the most extended methods for quantification of the $sp^2/sp^3$ ratio in carbonaceous materials. In the case of $CN_x$ solids, the relative contribution of the different hybridizations can be established independently for C and N atoms, allowing the identification of the different C-C and C-N bonding environments.

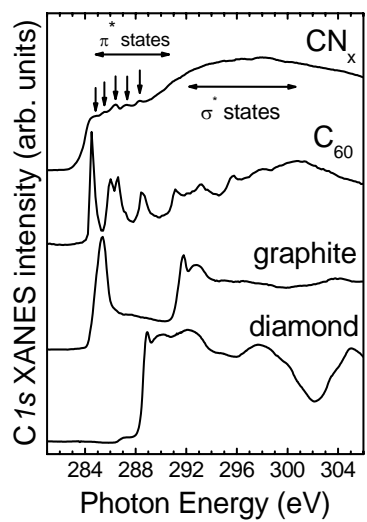

**Figure 6.** XANES C(1s) spectra from the carbon allotropes and a graphitic $CN_x$.

As an illustrative example to understand the XANES signal, Figure 6 shows the spectra from carbon allotropes (graphite, diamond, $C_{60}$) and from a typical graphitic $CN_x$ sample. The diamond spectrum shows a single absorption edge at 288 eV, corresponding to the threshold energy for transitions from the 1s level to the unoccupied $\sigma^*$ states characteristic of $\sigma$ bonds between $sp^3$-C atoms. The sharp peak at 289 eV is the $\sigma^*$ excitonic resonance, and the weak shoulder at lower energies correspond to transitions to C-H $\sigma^*$ states from the surface C-H termination. The graphite spectrum shows two absorption edges, a lower energy one at 284 eV corresponding to transitions to $\pi^*$ states, and the higher energy one at 291 eV related to the transitions to $\sigma^*$ states. These states are associated to the corresponding type of bonds, $\pi$ and $\sigma$, between $sp^2$-C atoms. The peaks at 285.4 and 291.8 eV correspond to the $\pi^*$ and $\sigma^*$ resonances, respectively. Regarding the $C_{60}$ spectrum, there is a series of discrete peaks that correspond to the transitions to empty molecular orbitals. Those appearing below 290.5 eV have a $\pi^*$ character, and those above this energy are of $\sigma^*$ type. Such discrete molecular orbital energies are lost in the case of extended systems like graphite. The spectrum of the $CN_x$ thin film displayed at the top of Fig. 6 possesses a graphitic structure since the spectral lineshape is similar to the graphite reference in the sense that it contains two separated absorption thresholds related to the $\pi^*$ and $\sigma^*$ states. The $\pi^*$ states lying below 290 eV show a series of peaks, with five being clearly resolved, as marked with small arrows in Fig. 6. However, a larger number of component curves may be present remaining unresolved. For different $CN_x$ samples, the relative intensity of the apparent peaks changes significantly, with a reduced intensity of the lower energy peaks in the samples with higher N intake (0.25>x>0.5). Despite several attempts from various authors to correlate each apparent peak with a specific bonding environment [37,39,40], this point remains unsolved.

For $CN_x$ solids, the study of N(1s) spectra is easier as compared to C(1s) because the π* peaks are better resolved [41]. A comparative study on many samples produced by different growth methods reveals that at least four components or $\pi^*$ peaks are present at the N(1s) edge [21,39]. These contributions are shown in panel (a) of Figure 7, with the peaks labelled from N1-N4. Peak N1 can appear in different energy positions within a narrow range (N1a-b). Although the two contributions have not been observed as separated resolved peaks, a careful examination of the N(1s) spectra indicates the presence of more than a single N1 peak. These N1 peaks are related to different pyridine-like environments, based on comparison with standards and theoretical simulations of the XANES lineshape [18]. The presence of several pyridine-like environments has been also observed by NMR [20] and can explain the asymmetry of the N1 peak due to the N1a-b contributions. From the simulation of XANES spectra and comparison with references, peak N2 has been assigned to nitrile groups (cyanogen-like) and peak N4 to three-fold coordinated N in the graphene sheet [18]. It should be noted that the peak position for substitutional sites in graphite or pyrrole-like environments yields very close, making difficult to differentiate between both situations [42]. The assignment of peaks N1, N2 and N4 to pyridine, nitrile and graphite-like environments, respectively, has been also stated by studying the angular dependence of the peak intensities in N implanted highly oriented pyrolitic graphite (HOPG) [43]. Peak N3 corresponds to molecular $N_2$ embedded in the film due to the ion assistance and gas trapped during the growth in a $N_2$–containing atmosphere, which can be released either at substrate or annealing temperatures above 400ºC [39]. A further support for this assignment is given in Fig. 7(b), where the oscillations corresponding to the XANES spectrum of $N_2$ molecules [44] is superimposed to that of a $CN_x$ film grown by DC-MS. A clear correspondence can be appreciated, with the $N_2$ oscillations superimposed to the spectral line signal. As a summary, Fig. 7(c) shows a sketch of the bonding model proposed to explain the XANES spectrum. Note that the incorporation of N in positions N1 and N2 involve the existence of C vacancies or holes in the graphene sheet.

There is additional evidence that the $CN_x$ structure is highly defective, with vacancy sites in the graphene planes playing a key role. A clear proof is the lack of a one-to-one correspondence between the C(1s) and N(1s) spectral lineshape, as evidenced in Figure 8. Here, the spectra from two $CN_x$ samples produced by different methods (labelled as A and B) are compared. Despite exhibiting a similar C(1s) spectrum, their N(1s) are very different. The opposite situation can be also observed for other samples, with similar N(1s) spectra and different C(1s) ones. This behaviour can be explained by the presence of vacancies, so that the nearest neighbour configuration around a C or N atoms depends not only on the number of N or C ligands, respectively, but also on the number of nearest neighbour vacancies.

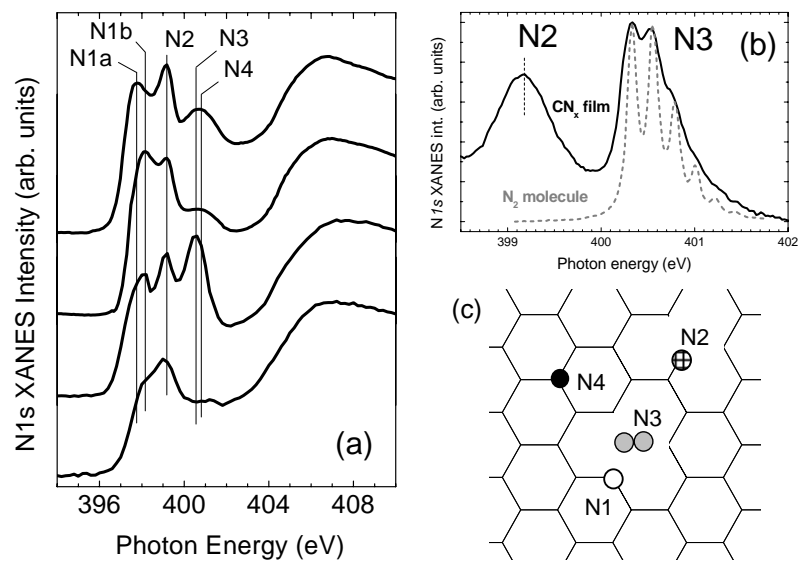

**Figure 7.** (a) XANES N(1s) spectra from selected $CN_x$ samples evidencing the presence of five component peaks. (b) Comparison of the spectral lineshape around peak N3 with that the XANES spectrum from $N_2$ molecules. (c) Bonding model explaining the different contribution in the XANES N(1s) spectra.

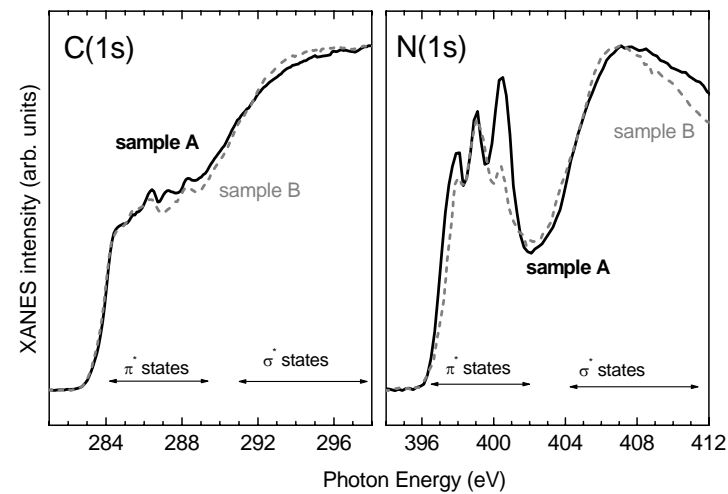

**Figure 8.** Comparison between the spectra of two $CN_x$ samples evidencing the lack of a one-to-one correspondence in the C(1s) and N(1s) curves.

In summary, the structure of $CN_x$ is based on defective graphene sheets containing vacancies, subtitutional three-fold coordinated nitrogen atoms, pyridine-like rings, and nitrile groups. As shown in detail afterwards, this bonding structure is found $CN_x$ films with an amorphous, graphitic and FL structures. Hence the formation of the FL arrangements should depend on the balance between the different bonding possibilities.

## 4.3. XANES spectral signatures related to FL microstructure formation

The FL structure in $CN_x$ is found for a certain window of growth temperatures and N content, with a well developed FL structure for samples near a $CN_{0.15}$ composition and grown at about 450ºC [22]. The N intake and the growth rate decrease with increasing temperatures, until a negligible growth rate is observed above 600ºC. However, the N content by itself is not a good parameter because growth at low or room temperature results in the incorporation of a significant fraction of embedded molecular $N_2$ that is not linked to the graphene-like structure and a detailed study of the different bonding environments is mandatory to correlate the microstructure evolution with the N incorporation. In order to address such correlation, we have studied $CN_x$ films grown by DC-MS with varying degree of FL character (in terms of plane alignment, extension and cross-linking). The FL degree is tuned by changing the $N_2$ content in the $Ar/N_2$ gas mixture at a growth temperature of 450°C or by keeping a low $N_2$ content (< 20%) and sweeping the growth temperature, as shown in Figure 9.

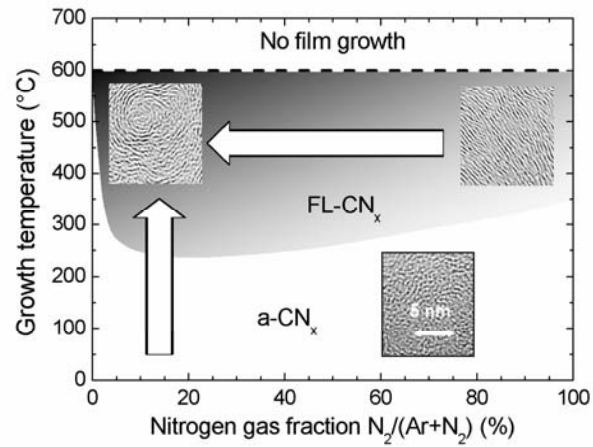

**Figure 9.** Sketch of the microstructures that can be obtained in $CN_x$ films grown by DC-MS as a function of growth temperature and $N_2$ content in the $Ar/N_2$ gas mixture. The grey zone corresponds to the parameter window for FL-$CN_x$. HRTEM images are only indicative. The arrows show the set of samples studied here with different degree of FL character, the direction following the increase of FL features.

Figure 10 shows the XANES spectra for a set of samples grown by DC-MS with 16% $N_2$ in the $N_2/Ar$ gas mixture at different substrate temperatures showing the evolution of the different bonding environments. The films grown for temperatures above 300°C presents a FL structure that is further promoted with increasing temperature. As a first conclusion, the spectra from $CN_x$ with and without the FL structure are very similar. As stated before, this indicates that the promotion of FL-$CN_x$ does not occur by the formation of a new or specific bonding environment but rather to the interplay, i.e. relative content and arrangement, of bonding environments commonly present in graphitic $CN_x$ structures. In this sense, it is clearly noted that the spectra of FL-$CN_x$ do not contain sharp peaks as occurs in the spectra of $C_{60}$ (see Fig. 6). These sharp features are related to the molecular orbitals of the $C_{60}$ structure. We must remind at this point that the FL microstructure in $CN_x$ thin films consists in a curvature and cross-linking of basal planes that extends to the whole film and is not forming nanometric size closed cages with molecular character.

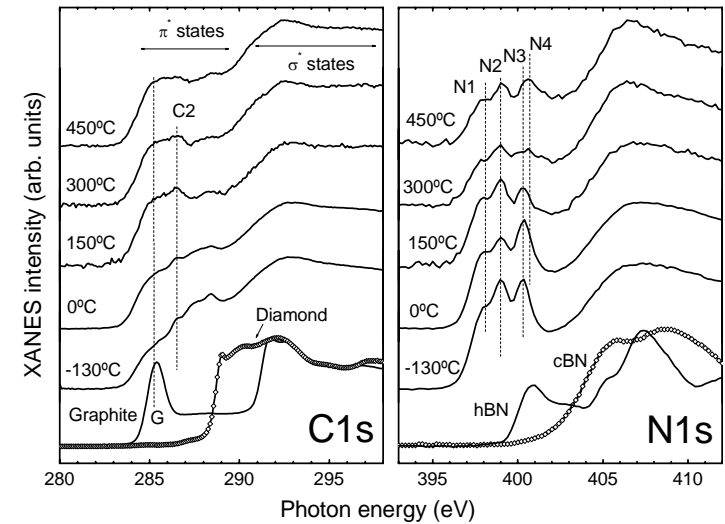

**Figure 10.** C(1s) and N(1s) XANES spectra for $CN_x$ films grown at different substrate temperatures. The spectra of graphite, diamond, cBN and hBN are shown as reference for $sp^2$ and $sp^3$ hybridizations.

Some trends can be extracted from Fig. 10. In the C(1s), the main observation is a intensity decrease of the $\pi^*$ states at higher energy with increasing temperature and an increase at lower energies (near the graphite reference peak G). At the N(1s) panel, there is a decrease of the N3 peak corresponding to molecular $N_2$ for temperatures up to 300ºC, with the subsequent appearance of the N4 peak for higher temperatures. Also, there is a decrease of the overall density of $\pi^*$ states in the N(1s) spectra, which is a general trend found in the FL-$CN_x$. The latter is made more evident from the spectra shown in Figure 11 corresponding to samples with a similar N content with FL (FL-$CN_x$) and planar (g-$CN_x$) arrangements. The density of $\pi^*$ states in the C(1s) spectra, computed as the area below the spectra between 282-289 eV normalized to the $\sigma^*$ states area between 292-299 eV, is almost the same in the two samples. However, the density of $\pi^*$ states at the N(1s) edge, computed as the area between 396-402 eV normalized to the $\sigma^*$ states area between 405-411 eV, is almost the half for FL-$CN_x$ as compared to planar graphitic structures (g-$CN_x$).

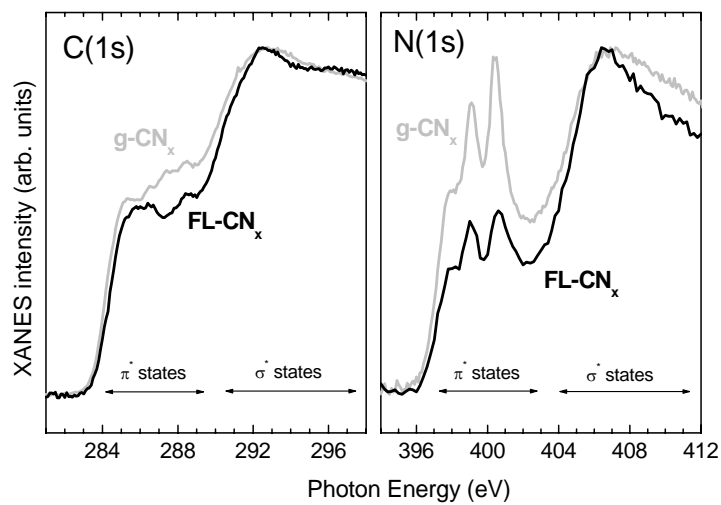

**Figure 11.** XANES spectra from a $CN_x$ films with FL and graphitic (planar) structures. There is a clear decrease of the intensity of N(1s) $\pi^*$ states in the FL case.

A more systematic study of the $\pi^*/\sigma^*$ correlation with the formation of FL arrangements can be obtained from the XANES spectra of Fig. 10. The results of the analysis are displayed in Figure 12, showing the evolution of the $\pi^*/\sigma^*$ intensity with growth temperature. We must recall that, in this case, the window for FL formation occurs at high growth temperature. Therefore, Fig. 12 corroborates that there is a clear correspondence between the formation of the FL structure and the decrease in the density of $\pi^*$ states occurring at the N atoms. Another striking results is that the density of $\pi^*$ states at the C atoms remains essentially constant, i.e. there is no appreciable $sp^3$ C creation upon FL formation.

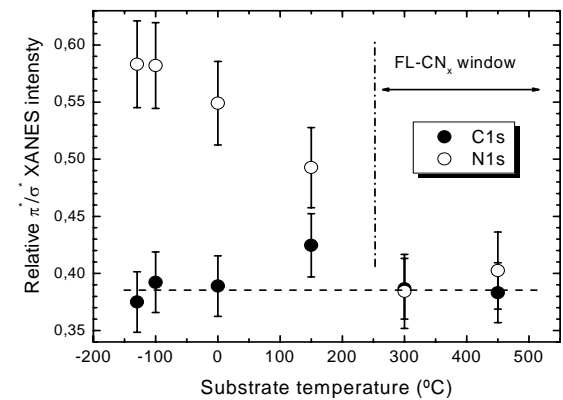

**Figure 12.** Relative $\pi^*/\sigma^*$ XANES intensity at the C(1s) and N(1s) absorption edges for the spectra shown in Fig. D. The evolution towards the FL structure implies a decrease in the contribution of $\pi$ electrons from N atoms, while it is kept constant for C atoms.

Apart from the growth temperature, the other important parameter in the microstructure evolution is the N content, $x$, in the $CN_x$ films. Figure 13 shows the XANES spectra from a series of samples grown by MS at 450ºC with different proportions of $N_2$/Ar in the processing gas, namely 100%, 50%, 16% and 10% of $N_2$ yielding the $x$ values of 0.24, 0.2, 0.17 and 0.1, respectively. HRTEM images only show a clear FL structure for the two samples with the lower $N_2$ contents in the gas mixture. By direct observation of the XANES spectra it is clear that the samples exhibiting the FL structure are those with a dominant peak N4, which corresponds to substitutional N three-fold coordinated to carbon in the graphene-like sheets. However, for samples lacking the FL structure, the dominant peak is N2 related to nitrile groups. Coming back to the samples grown at different temperatures shown in Fig. 10, the same observation can be made: the sample with dominant FL structure (i.e. the one grown at 450ºC) has a dominance of peak N4 in the N(1s) spectrum.

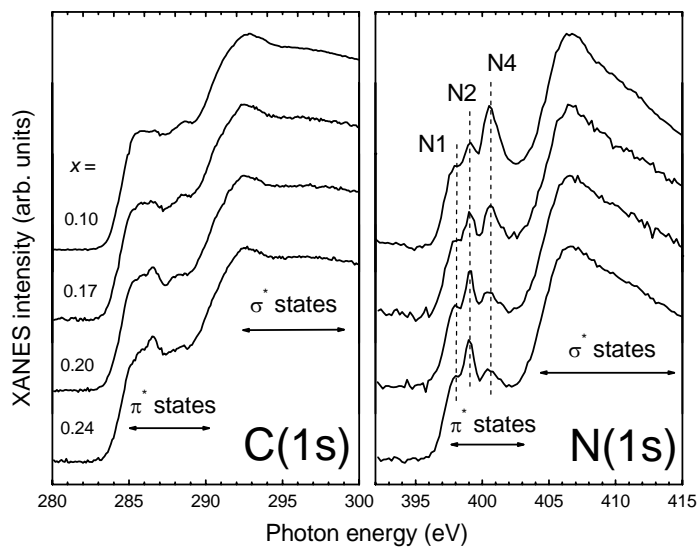

**Figure 13.** C(1s) and N(1s) XANES spectra for $CN_x$ films grown with different $Ar/N_2$ gas mixtures at a substrate temperature of 723 K, yielding the different nitrogen contents x indicated.

In summary, the two spectral features related to the FL structure are: (i) a dominant peak N4 in the N(1s) XANES, and (ii) an overall decrease of the intensity of $\pi^*$ states. Both observations support that the formation of the FL structure require a dominant number of substitutional N atoms incorporated with three-fold coordination in the graphene sheet.

### 4.3. Origin of curvature and nature of cross-linking sites

In this section we discuss the different bonding models that have been proposed to explain the formation of a FL structure, taking into account that they must explain the curvature of the graphene-like sheets and their cross-linking. In analogy with to $C_{60}$ molecules, the bending of the basal planes has been initially attributed to formation of odd-member rings induced by the presence of N [3]. This assumption was supported by theoretical calculations showing that N incorporation can simultaneously reduce the energy required to create pentagon defects and favour the formation of reactive C-$sp^3$ adjacent sites [16]. Accordingly, the formation of C-$sp^3$ sites was considered as the cross-linking path in FL-$CN_x$ [13,16]. Contrary to this model, the spectroscopic measurements on $CN_x$ films by NMR [20,21] and the XANES results of Fig. 12, have shown that the bonding structure is composed of ~100% $sp^2$-C, hence, neglecting the $sp^3$-C contribution. Other plausible folding mechanisms consider the corrugation induced by non-planar N atoms in graphitic substitutional sites [17].

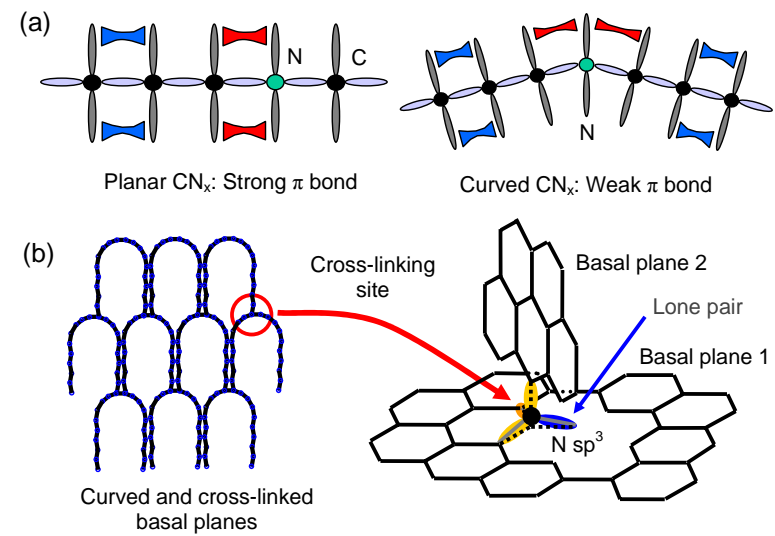

**Figure 14.** Geometrical model illustrating the role of N in the (a) reduction of π bonding in curved $CN_x$ structures and (b) cross-linking of two perpendicular graphene-like planes through a non-planar N atom.

The trends in the $\pi^*/\sigma^*$ signal displayed in Fig. 12 provide important information about changes in the localization of π electrons. First, it shows the relevant role of N in curved graphitic $CN_x$ structures due to the fact that the decrease in the contribution of N to π bonds is not accompanied by a parallel decrease in the C contribution. For planar atomic arrangements, C and N atoms form strong π bonds with significant participation to of p-electrons from the respective atoms. Therefore, the π electron cloud is delocalized over extended areas including both C and N atoms. In the case of $CN_x$ corrugated or FL arrangements, there is a disruption of the conjugation of the π electrons and the π electron cloud is more localized in regions composed only of C atoms. This geometrical configuration is illustrated in Figure 14(a), showing a reduced contribution of N atoms to π bonding. The different localization of π electrons in planar and corrugated $CN_x$ must also affect the elastic properties of the graphene sheets and, therefore, could explain the outstanding mechanical properties of FL-$CN_x$. Following the approach of non equivalent sites for C and N in the graphitic network, it has also been proposed that non-planar N atoms can additionally act as the cross-linking sites between adjacent sheets [23]. A sketch of this geometrical situation is shown in Figure 14(b), with two perpendicular planes being cross-linked through a non-planar N($-sp^3$) atom. Note that the N atom is located adjacent to a C vacancy, which seems to be requisite for many of the proposed models of curvature and cross-linking.

There is another model considering a mechanism for curvature that stems from the high number of vacancies present in the graphene sheets upon N incorporation. Here, the presence of holes in the structure permits the presence of pentagonal rings without the need of a neighbour $sp^3$-C atom. Following this argument, it has been proposed that cross linking occurs at the dangling bond (two-fold coordinated) sites on the perimeter of a vacancy, a situation especially favourable when the pentagon is placed next to a single-atom vacancy because it will have a twofold coordinated vertex [20]. At present, a clear identification of the folding mechanism and cross-linking of $CN_x$ planes remains unclear.

## 5. ON THE MECHANISMS OF FL-$CN_x$ FORMATION

### 5.1 Comparison of different nitrogen incorporation routes

In this section, the microstructure of $CN_x$ films grown under different growth methods that mimic partially the growth by MS is studied. Namely, ion beam sputtering (IBS), dual ion beam deposition (DIBD), ion-beam assisted evaporation (IBAE) and magnetron sputtering (MS). The principles of these methods, together with that of MS, are illustrated in Figure 15.

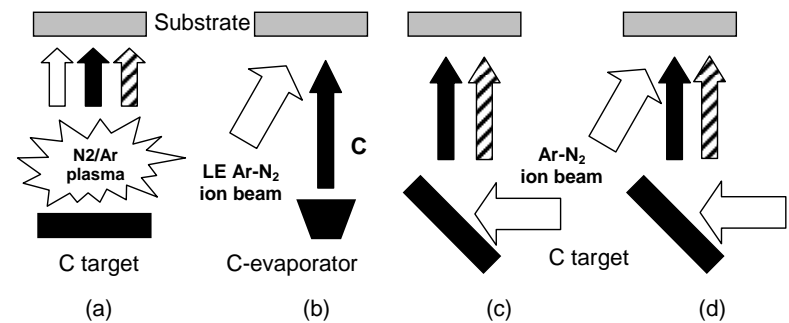

**Figure 15.** Different growth methods compared in this study in terms of ion bombardment (white arrows), C fluxes (black arrow) and polyatomic $C_xN_y$ species (striped arrow): (a) MS, (b) IBAE, (c) IBS and (d) DIBD.

Each one of the methods presented (MS, IBAE, IBS and DIBD) implies different N incorporation routes. In the case of IBAE, the sole source of C comes from the thermal evaporation of graphite lumps with an electron evaporator whereas N is incorporated by low-energy ion assistance (<100 eV) with a mixture of $N_2^+$ and $N^+$ ions extracted from an End-Hall grid-less ion gun [45]. In the case of sputtering from graphite targets by $N_2$/Ar ions of a few hundreds of eV (MS, IBS, DIBD), C and N atoms can be supplied trough atomic and polyatomic $C_xN_y$ species, the latter chemically or physically emitted from the previously nitrided target [31]. The emitted particles via physical sputtering have higher energies (in the hyperthermal range up to a few tens of eV) than in the case of thermal evaporation (in the range of ~0.1 eV) [46,47]. The additional application of Ar/$N_2$ ion assistance in DIBD (~100 eV) and MS (25-40 eV) represents, together with the incorporation of $C_xN_y$ species, an additional source of N during the film growth. The growth conditions obtained by DIBD are the most similar to MS with the advantage that the weight of the different effects can be tuned. Finally, during sputtering of the target reflected ions may also be produced with energies up the initial value of the projectile (several hundreds of eV) [48]. The high energy of these ions could play a significant role in the growth, however, their effect can be neglected according to calculations done with the SRIM code [49].

## 5.2. N incorporation in the films from the sputtering target

In this section we present the results obtained by IBS of a C target with a $N_2$/Ar ion beam. In this way, we can differentiate the N incorporation from the sputtering target from the additional influence of $N^+$ ion assistance that can not be suppressed in the case of MS. This incorporation has to be understood as a competitive process from N sources of N sputtered atoms and $C_xN_y$ precursors emitted from the nitrided target surface. Another plausible incorporation of N comes from trapping of the residual $N_2$ gas.

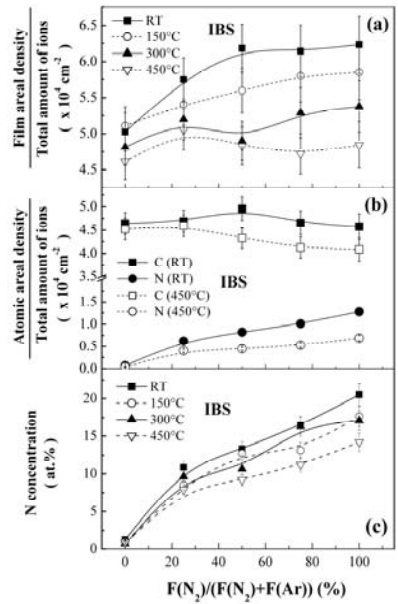

**Figure 16.** (a) Ratio between the film areal density and the total amount of sputtering ions delivered by the ion source; (b) partial contribution from C and N atoms to the film areal density; and (c) nitrogen concentration as a function of the $N_2$ content in the $N_2$/Ar sputtering gas mixture at different substrate temperatures (from Ref. 50). The lines are only guide-to-the-eye.

The interplay of the different incorporation mechanisms can be deduced from Fig. 16(a), where the total film areal density, partial areal densities of C and N atoms and the N atomic ratio, as measured by elastic recoil detection analysis (ERDA), is presented as a function of the sputtering Ar/$N_2$ gas mixture. It can be seen that the total amount of deposited atoms at low growth temperatures increases with the availability of $N_2$ up to a flow of 50% and then saturates. The origin of the initial increase of deposited atoms can be extracted from Fig. 16(b), where the normalized amounts of incorporated C and N atoms for the samples deposited at RT and ~450°C. The C incorporation does not change significantly with the sputtering beam composition at RT, while the N incorporation scales with the $N_2$/Ar gas ratio. This results in the net increase of the total number of incorporated atoms. At higher temperatures, this tendency is preserved except that for $N_2$ mass flow contents higher than ~50% there is a slight decrease in the C incorporation while the number of incorporated N atoms continues to increase, which result in almost constant normalized film areal density. Fig. 16(c) shows that a significant amount of N can be incorporated into the films just by sputtering the carbon target (~10-20 at.%) with an Ar/$N_2$ beam. The maximum N concentration obtained by this method (~20 at. %) is comparable with that obtained in $CN_x$ films grown by MS or filtered cathodic vacuum arc deposition, which it is in the range between 35 at.% at temperatures below 150°C [25,51] and 17 at.% at 550°C-600°C [13,52,53].

Due to N ion bombardment of the C target, N is constantly implanted into the near surface layer and sputtered by subsequently incoming ions. The rate between sputtering and N implantation determines the N concentration on the target surface. In this context, sputtering of the C target with N ions has to be considered as a hybrid process of chemical and physical sputtering of atomic and $C_xN_y$ species (x, y ≤ 2) from the target surface [31]. These emitted single- and multi-atomic species impinge on the substrate and are adsorbed. SRIM calculations give similar sputtering yields for N and Ar ions [50], thus, similar film areal densities can be expected for different sputtering beam compositions. However, this contradicts the observed larger slope of N concentration when the $N_2$ content in the sputtering gas increases from 0 to ~25% than that for larger $N_2$ mass flow ratios (Fig. 16(c)), and the observed concomitant increase in the film areal density at RT with $N_2$ content in the sputtering beam (Fig. 16(a)). This indicates that with $N_2$ addition to the $N_2$/Ar sputtering beam, another sputtering mechanism starts to take place simultaneously with physical sputtering produced with pure Ar sputtering beam. This can be related to the changes in surface binding energy with the changes in target surface composition and chemical sputtering effect [54].

Due to the similar mass of C and N atoms, their physical sputtering yield will be close. This implies that the N concentration in the film has to be proportional to the N concentration in the target surface, which should be proportional to the composition of the sputtering beam. Following this, if there was a saturation of the N concentration at the target surface for certain beam composition, this would be reflected in the film composition. On the other hand, if the preformed radicals were sputtered preferentially in relation to atomic species, then the film composition would reflect not the target composition but the mean composition of $C_xN_y$ species. In this case the N concentration in the film can saturate while the N content at the target surface will still increase with increasing nitrogen content in the sputtering beam. However, no N saturation at increasing $N_2$ ratio in the sputtering beam is observed, contrary to the observations with MS [29,55]. The saturation behavior in MS was interpreted as a result of preformed $CN_x$ species coming from the graphite target which are the main source of incorporated nitrogen while nitrogen from the reactive plasma does not serve as a direct nitrogen source [29,55]. This should also apply to the present experiment where the main N source is the graphite target being irradiated with $Ar/N_2$ ions. However, the observed concomitant increase in N concentration with the $N_2$ ratio in the sputtering beam implies that a significant amount of film forming species constitutes of physically sputtered atomic species. The main difference between MS and IBS is that for the latter technique there is no plasma assisting the film growth. Therefore, the saturation of the N concentration during MS has to be related to the N transfer due to surface-plasma interaction and with some dynamic equilibrium between N gain and N loss and not with the preformed species.

The main mechanisms responsible for the material loss from the growing film can be classified in the following way: 1) temperature reduced incorporation of the volatile molecular species originating from the residual gas or the target; 2) the temperature induced/enhanced formation and desorption of volatile molecules (mainly, $C_xN_y$ and $N_2$) on the surface and in the subsurface layers of the growing film. Both mechanisms can be expected to act simultaneously. The presence of hydrogen found in the films [50] indicates about the adsorption of molecular species and their trapping due to incoming particles, because hydrogen is present mainly in the form of hydrocarbons and water. The H content decreases concomitantly with temperature which indicates about reduced adsorption of such molecular species on the surface of the growing film. This concerns also the $N_2$ molecules originating from residual gas or volatile $C_xN_y$ compounds originating from the target surface. This correlates well with the results presented in Fig. 14, where the considerable decrease in C areal density is followed by the decrease in N areal density when the temperature increases from RT to 450°C (see Fig. 14(b)). However, it also correlates with the nitrogen induced chemical sputtering by forming volatile compounds with C, mainly $C_2N_2$. Their formation rate will depend on the N availability on the surface and subsurface layers and will result in higher sputtering rate for higher N concentrations on the surface of the growing film as it is in Fig. Fig. 14(b). The surface reaction rate is proportional to the mobility of the atomic species, and thus to the substrate temperature. Besides, the computer simulations of the growth of $CN_x$ films indicated that for N loss through the chemically enhanced mechanism to proceed, it is necessary that at least one of the reacting atoms has been set into motion ballistically [54]. This is in agreement with the film growth conditions using IBS where the incoming particles exhibit some hyperthermal energies.

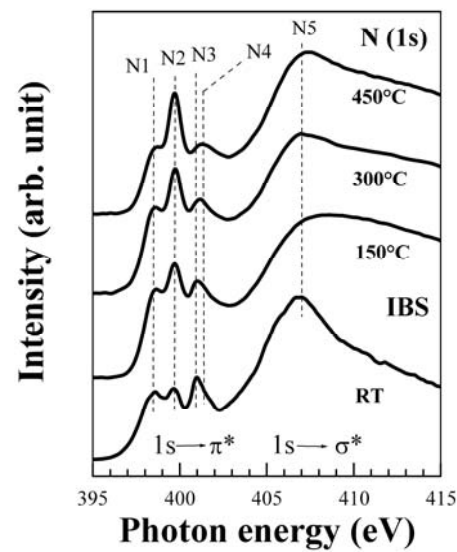

**Figure 17.** N (1$s$) XANES spectra of $CN_x$ films grown by IBS with pure nitrogen beam at RT, 150°C, 300°C and 450°C (from Ref. 50).

### 5.3. Bonding structure of $CN_x$ films grown by IBS

The local nitrogen chemical bonding structure of films grown by IBS is shown in Figure 17. The assignment of the different spectral peaks (N1-N4) was introduced in Section IV. It can be seen that the contribution of the nitrile-like N (peak N2) increases concomitantly with temperature. This type of terminating bonding hinders the formation of the extended basal graphene sheets. As the total number of incorporated N atoms decreases with increasing temperature, this indicates that the nitrile-like configuration is the most stable against the temperature induced/enhanced desorption process. On the other hand, the contribution coming from pyridine-like environments (N1), molecular nitrogen (N3) and three-folded N (peak N4) decreases. The apparent shift of peak N3 towards higher photon energies with temperature should be interpreted as the decrease of peak N3 and the emerging of peak N4 due to the desorption of embedded $N_2$ molecules and the promotion of three-fold N. This observations are commonly reported in graphite-like $CN_x$ structures [21,23,37,39].

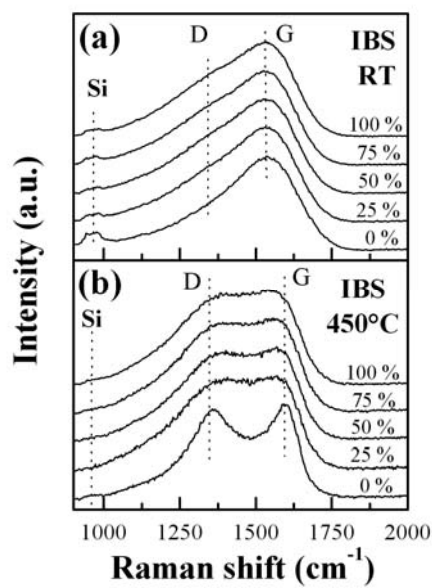

**Figure 18.** Normalized Raman spectra of films grown by IBS at RT (a) and 450°C (b) for different $N_2$ content in the sputtering beam (from Ref. 56).

Further information about the bonding structure and microstructure of IBS $CN_x$ can be obtained by Raman spectroscopy [56], which is sensitive to middle range order in carbon-based materials. Figure 18 present typical Raman spectra in the wave number region of 900-2000 cm$^{-1}$ for IBS films grown with different $N_2$/Ar ratios at RT (panel a) and 450°C (panel b). Raman spectra of C and $CN_x$ films in this wave number region exhibit usually two main peaks positioned at ~1350 and 1560-1590 cm$^{-1}$ which are denoted conventionally by D and G, respectively. The G ('graphite') peak is due to in-plane bond stretching vibrations. This mode is a single resonance and can be present in aromatic clusters as well as in chain structures [57]. The D ('disorder') mode is a double resonance [58] and is associated with breathing vibrations of aromatic rings. The D peak is not present in the spectrum of single crystalline graphite or in completely amorphous C films. The intensity of this peak is due to two concurrent factors: (i) the increase of the probability of a double resonance event when an excited electron is scattered by a phonon and a defect (i.e. grain boundary) when the size of the graphitic grains decreases [58], and (ii) the presence of ring structures [57].

The origin of the D and G peaks allows estimating the six-fold ring cluster size $L_a$ by comparing the intensity of the G peak (vibrations present in the $sp^2$ phase in olefinic and ring structures) with that of the D mode (vibrations only coming from six-fold rings). Following the three stage model proposed by Ferrari *et al.* [57], the reduction of the grain size in crystalline graphite will induce the appearance of the D peak, whose intensity will increase as the grain size decreases. In this regime the Tuinstra-Koenig relation is valid [59]:

$$L_a = 44 \times (I_D/I_G)^{-1} \text{ (Å)} \qquad (1)$$

for the excitation laser wavelength of 514.5 nm. This expression is valid down to critical cluster size of 20 Å, below which the second mechanism starts to act and the cluster size is given by [57]:

$$L_a = \left(\frac{I_D/I_G}{0.0055}\right)^{1/2} \text{ (Å)} \qquad (2)$$

In this regime, six-fold ring clusters become smaller and more disordered. When the disorder increases, the $sp^2$ phase changes from ring to chain structures, while at the final stage when $sp^3$ bonds are introduced the $sp^2$ bonds become strongly localized.

D and G peaks can be also observed for the majority of the spectra grown by IBS (Fig. 18) indicating a presence of the $sp^2$ phase, and for some of the spectra a small peak at $\sim 960$ cm$^{-1}$ of the underlying Si substrate could be also seen. For the IBS films grown at RT, the D peak contribution is very low and can be hardly resolved in the spectra. Two main observations are the slight increase in the intensity of the D peak with N incorporation and temperature (see Fig. 18). The former indicates N induced clustering of the $sp^2$ phase into ring structures, while the second points to a temperature induced graphitization. All the spectra lack sharp features, except the C film (grown by pure Ar bombardment) grown at 450°C. However, these sharp features of the D and G peaks are not present under the same growth conditions when N is incorporated, indicating increased disorder.

Quantitative information can be extracted by applying certain fitting procedure to the observed experimental spectra. The combination of a linear background, a symmetric Lorentzian and and asymmetric Breit-Wigner-Fano (BWF) shape to fit the D and G peaks, respectively, was found to be a good compromise between the fitting quality and the capability to fit all spectra. The BWF line shape is described by the following expression [57,60]:

$$I(k) = \frac{I_0 [1 + 2(k-k_0)/q\Gamma]^2}{1 + [2(k-k_0)/\Gamma]^2}, \quad (3)$$

with the maximum positioned at [57]

$$k_{max} = k_0 + \Gamma/2q, \quad (4)$$

where $I(k)$ is the intensity as a function of wave number $k$, $I_0$ is the peak intensity, $k_0$, $k_{max}$ and $\Gamma$ are the mean peak position, the peak position at the maximum, and the full width at half maximum (FWHM), respectively, and $q$ is BWF coupling coefficient. In the following the position of the G peak will be defined by $k_{max}$. In the limit $q \to 0$, the Lorentzian line shape is reproduced. Generally, the G peak position reflects the bonding strength, and $\Gamma_G$ reflects bond length and angle distortion [14,60,61]. In addition, implantation and annealing studies of graphite and glassy carbon [60,62] have established that in the limit $\Gamma_G/q \to 0$ the in-plane graphene ordering is complete, and three-dimensional ordering begins. Following this, $\Gamma_G/q$ will be used to quantify the in-plane graphitic order.

Raman spectra fitting results of the CN$_x$ films samples grown by IBS are plotted in Figure 19. From the dependence of the G peak positions and the G width $\Gamma_G$ on the N atomic ratio of the films, two different growth regimes can be identified: below and above $\sim 150$°C. At RT the position of the G peak slightly increases as the N atomic ratio in the film increases, while at higher temperatures it decreases. For a given N atomic ratio the G peak shifts towards larger values when the temperature increases, and this shift is larger for lower N atomic ratios. Opposite trends can be seen for the $\Gamma_G$ dependence.

Additional information can be obtained by plotting the coupling parameter of the G peak $\Gamma_G/q$ and the D peak width $\Gamma_D$ versus the N atomic ratio. Both parameters exhibit very similar behavior, reflecting the changes in the ring structures. For pure C films, $\Gamma_G/q$ approaches zero when the temperature increases, which means that graphene plane ordering increases and approaches the stage when three-dimensional ordering begins [62]. However, when some N is introduced into the films it increases significantly the disorder in the ring structures, which is more pronounced for higher temperatures. Increasing the N atomic ratio above $\sim 10$ at.\% does not induce any significant changes in both $\Gamma_G/q$ and $\Gamma_D$. At the same time, for a given composition, $\Gamma_G/q$ and $\Gamma_D$ do not vary significantly with temperature, and this variation is lower for higher N atomic ratios. For the IBS films grown at RT, $\Gamma_G/q$ is approximately equal to $\sim 80$-$85$ cm$^{-1}$ and does not depend on the N content.

The evolution of the six-fold ring cluster size can be traced by plotting the cluster diameter $L_a$ versus N atomic ratio for different temperatures (see Fig. 19(e)). One can observe that at RT, $L_a$ increases with N incorporation. For 150 and 300°C, $L_a$ is higher than that obtained at RT and does not depend on the N content while at 450°C N incorporation prevents the extension of graphene planes. The last observation indicates that N incorporated into the C matrix during IBS enters into bonding configurations which prevent the extension of the graphene planes. According to this, a high quantity of terminating pyridine-like or nitrile-like configurations could be expected, which is in agreement with XANES observations shown previously.

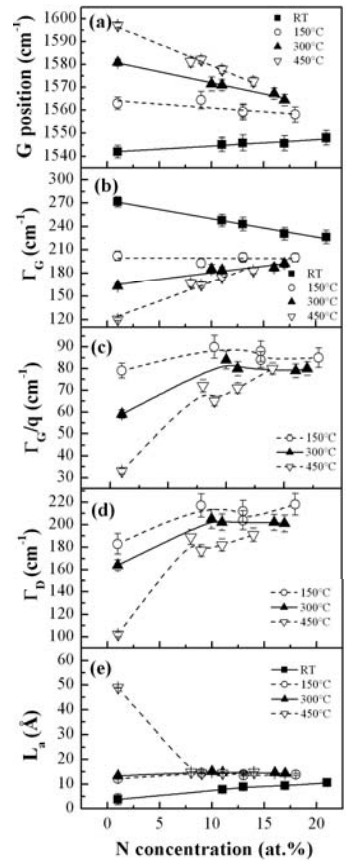

**Figure 19.** Results from the fitting of the Raman spectra of C and $CN_x$ films grown by IBS as a function of the N content (from Ref. 56.): (a) position of the G peak; (b) and (d) widths of the G ($\Gamma_G$) and D ($\Gamma_D$) peaks, respectively; (c) coupling parameter of the G peak; and (e) cluster diameter, $L_a$. The lines are only guide to the eye and the error bars represent the 95\% confidence limits of the fittings.

The results above indicate that N plays a double role creating ordering or disorder depending on the growth temperature. At RT, an increasing N content in the films induces a decrease in the bond length and angle distortion (see also Refs. 14, 51 and 63). This is related to 2D $sp^2$ ordering and six-fold ring clustering resulting in higher $L_a$. In this regime N incorporation is equivalent to an increase in temperature. When the films are grown above RT, $L_a$ increases compared to the films grown at RT. However, $L_a$ is nearly independent of both N content and substrate temperature in the 150-300°C range. At 450°C N incorporation in the C network results in a sharp drop in $L_a$ yielding the $L_a$ values similar to that obtained at 150-300°C. The following N incorporation does not result in any significant change of $L_a$. The fact that $L_a$ decreases with low N addition indicates that N is partially positioned at cluster edges and suppresses the temperature induced cluster growth. This is probably due to formation of pyridine-like $sp^2$ or nitrile-like $sp^1$ bonds. $L_a$ becomes relatively low (~13-15 Å), so that relatively large part of the cluster atoms is assumed to stay at the edges rather than inside the clusters. This means that a low fraction of N atoms in the terminating configurations can inhibit the cluster growth. However, the drop in $L_a$ with N incorporation at ~450°C is accompanied by a significant change in $\Gamma_G/q$. This may indicate that N substitutes C in six-fold rings, thus introducing high disorder inside the clusters. Additional N incorporation does not change $\Gamma_G/q$ significantly. This indicates, that the following N addition does not change the disorder inside the cluster, thus it can again be deduced that N is accumulating on the perimeter of the aromatic clusters. This corresponds to XANES observations of IBS $CN_x$ films where N was found in the one-, two- and three-fold bonding configurations [50].

### 5.4. Effect of simultaneous ion bombardment on the bonding structure

The influence of assisting ion bombardment is very complex. Here, we can test its influence by adding a secondary ion beam to the IBS process. This method is normally found in the Literature as dual ion beam deposition (DIBD) (see Fig. 15). First, when N ions are incorporated in the beam it acts as an additional N source. In fact, the N concentration of the $CN_x$ samples grown by DIBD is almost by a factor 2 higher than that obtained by IBS with a pure $N_2$ beam [50]. In addition, the N content obtained by DIBD with N ion assistance and the graphite target sputtered with a $N_2$/Ar beam is twice higher than that obtained with a N ion beam was used for assistance and a pure Ar beam was employed for carbon target sputtering [64]. Moreover, the nitrogen concentration of DIBD sample is very similar or higher to those obtained with MS at similar temperatures [50]. Then, it follows that the N incorporation from the assisting beam is at the same level as the N source from the nitride graphite target.

Low-energy assisting ion bombardment (<100 eV) can also activate or enhance surface processes such as diffusion but, it can also introduce some atomic displacements at sufficient high energy transfer. The displacement energy is of the order of the order of ~40 eV for graphite and diamond [65,66]. These values refer to bulk displacement energies while those near the surface can be significantly lower due to lower coordination number (lower binding energy) and lower displacement radius. If the ion energy is lower than the threshold for atomic displacements, it can still induce bond-breaking of pre-existing surface configurations [67], thus inducing chemical reactions which may also involve the incoming particle [68]. The energy threshold for bond breaking in C-C system is of the order of several eV [69], thus, having more probability to occur than the displacement event. In addition, molecular dynamics simulations show that surface processes dominate when the energy of impinging atoms is <30 eV, while subsurface processes govern the structure formation for higher energies [67].

Sputtering-based methods such as IBS or MS can be considered energetic growth processes themselves due to the energy of the emitted particles (mostly neutrals) in the hyperthermal range (from a few eV up to tens of eV) [46] that may activate or enhance surface processes. Consequently, hyperthermal atoms in the range up to ~20 eV [47,48] originated from the target due to sputtering provide a non negligible energy input on the surface of the growing film, which results in the formation of a compact structure at low deposition temperatures [56,70]. In addition, it can be expected that large amount bond are broken at the surface under the impingement of emitted particles from the target, as well as from assisting ions. A majority of preformed $C_xN_y$ (x,y≤2) radicals coming from the sputtered target an be expected to be affected by bombardment of energetic target species and from the assisting plasma/ion beam, making them available for chemical reactions, so that new atomic configurations can be formed under the influence of atomic mobility and ion bombardment [32]. It can be related to the fact that different atomic configurations have different damage thresholds [71] that would promote the preferential evolution of the phase with a higher damage threshold.

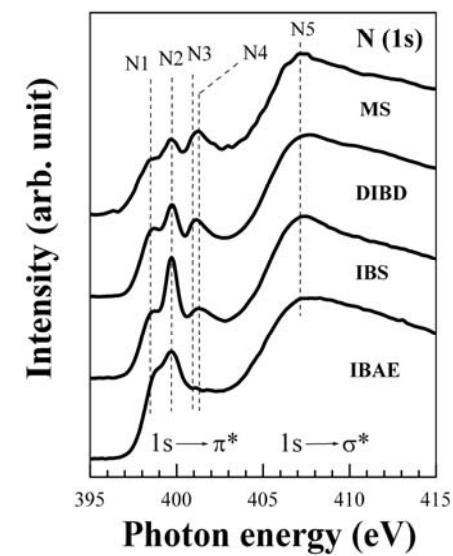

**Figure 20**. N (1$s$) XANES spectra of $CN_x$ films grown at ~450°C by IBAE, IBS, DIBD and MS (from Ref. 50).

Figure 20 compares the N(1$s$) XANES spectrum of $CN_x$ films grown by IBS, DIBD, IBAE and MS. The MS sample has pronounced FL features, which are identified in the XANES spectrum by the promotion of three-fold nitrogen (peak N4) over pyridine-like and nitrile-like bonding environments (peaks N1 and N2) (see section IV). The main difference between MS and IBS is the absence of plasma assistance during the film growth, which seems to be crucial for the promotion of N incorporation into graphite-like sites (peak N4). This statement is in agreement with the observation that the introduction of the low energy assisting ion beam during IBS in the DIBD configuration significantly reduces the fraction of N atoms in nitrile-like configuration while increasing the nitrogen fraction at graphite substitutional sites. This agrees with the expectation that the displacement or bond breaking threshold is lower for C≡N groups than for $sp^2$ C and N atoms inside or on the edges of aromatic clusters and, hence, these groups may be preferentially removed under low-energy ion bombardment. The removal of C≡N groups at relative low ion damage may explain the higher extension of grapheme sheets in MS thin films. This mechanism may also partially explain the formation of the texture with standing graphitic basal planes during MS growth of FL-$CN_x$ films [18] since the compressive stress usually present in such films cannot explain this phenomenon [72]. However, if the degree of ion bombardment is high enough, then formation of any ordered atomic configuration may be stopped, which which can lead to a complete amorphization of the growing film [7] and nitrogen incorporates only at sites with low coordination number as those observed in the films grown by IBAE [45].

Apart from changes in the local bonding configurations, ion bombardment can introduce significant material loss due to the (sub)surface displacements and bond breaking events [45,56] leading in some cases to a complete re-sputtering even at low ion energies (~25 eV) [25]. The material loss is observed with increasing the degree of ion bombardment or the substrate temperature, while N/C ratio in the film remains approximately constant [45,56]. The former indicates that an excess in the N ion to atom ratio results in a loss of film forming material (C and N) and that C and N losses are both correlated. This can be explained by ion enhanced/induced formation of volatile CN compounds via bond splitting and following chemical reactions [27,37,54,73,74,75].

### 5.5. Microstructural studies of $CN_x$ grown by different methods

In order to correlate the different growth methods (N-incorporation routes) and the obtained bonding structures with the microstructural evolution, HRTEM investigations were carried out in representative samples (for more details, see Refs. 45 and 76). In this case, $CN_x$ films grown by IBAE and IBS with a similar content as that obtained in FL-$CN_x$ by DC-MS have been studied. Also, a pure C film grown by IBS and a $CN_x$ sample grown by DIBD were investigated to discern the effect of N incorporation and ion bombardment effects. The HRTEM micrographs for the samples grown by IBS and DIBD are shown in Figure 21.

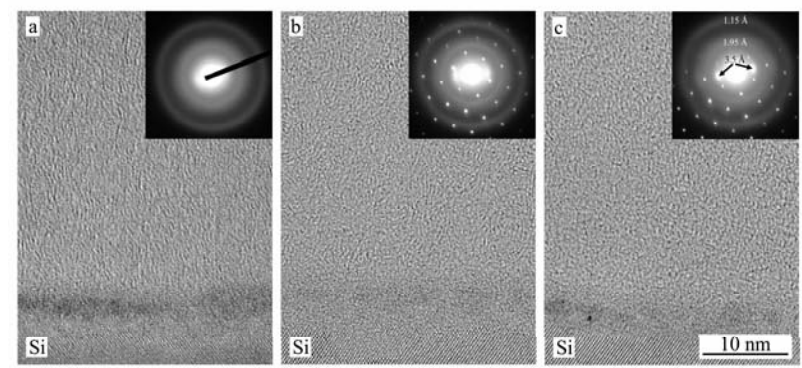

**Figure 21.** Cross-sectional HRTEM images and SAED patterns for (a) C and (b) $CN_x$ films grown by IBS at 723 K. Panel (c) shows a $CN_x$ film grown by DIBD at 723 K under concurrent 100 eV nitrogen ion assistance (from Ref. [76]).

FL arrangements were not found in IBAE $CN_x$, which means that low-energy N ion bombardment together with an atomic C flux is not the only requisite for FL formation [45]. On the contrary, the image of a C film grown by IBS (Fig. 21(b)) shows a highly textured microstructure consisting of predominantly standing basal planes. The introduction of N by IBS (Fig. 21(b)) reduces the texture but promotes curvature in the basal planes, in agreement with the hypothesis that N is crucial for the evolution of heavily bent and frequently cross-linked basal planes at much lower energies as compared to pure C films. The achievement of FL arrangement in the case of IBS and not in IBAE indicates that the sputtering target plays a significant role in the evolution of the microstructure and supports the eventual participation of preformed $C_xN_y$. Finally, Fig. 21(c) indicates that although some microstructure remains, the relatively high-energy involved inhibits the evolution of extended FL domains (lower energies could not be used due to experimental constraints but future experiments are in progress to reduce the energy of the assistance ions and, in this way, check the promotion of FL arrangements by ion assistance).

The SAED patterns displayed as insets in Fig. 21 give further indications of the film microstructure since projection artifacts do not affect the characteristic lattice spacing in the diffraction pattern and information on the degree of ordering can be derived from the brightness and width of the diffraction pattern [34]. These patterns show diffuse features indicating the quasi-disordered structure of the film corresponding to lattice spacing of ~1.15, 1.95 and 3.5 Å (see labels in the inset of Fig. 21(c)). In the insets of panels (b) and (c) in Fig. 21, the dot diffraction pattern of the crystalline Si(100) substrate is also superimposed to indicate the relation of the texture of FL features to the substrate normal. The rings at ~1.15 and ~1.95 Å coincide with those observed for amorphous allotropes of C and $CN_x$ films while the innermost ring at 3.5 Å match with the inter-plane separation of hexagonal basal planes in graphite (0002). The patterns show a broad feature at 3.5 Å indicating the presence of defective graphitic basal planes. The appearance of a broad arc is due to the preferential orientation of the basal planes perpendicular to the surface normal. Well-structured FL-$CN_x$ presents a similar diffuse pattern but with more distinguishable features and an additional ring at 1.75 Å [34]. The intensity of this ring correlates with that at 3.5 Å and, therefore, it is attributed to the inter-plane distance in graphite (0004). The presence and definition of the 3.5 and 1.75 Å rings have been correlated with the evolution of the FL structure, as signatures of the short/medium range graphitic order [34]. In the case of IBS films, the characteristic 3.5 Å spacing shows similar structured films as those grown by DC-MS, but the diffuse ring at 1.75 Å is less defined. This feature is only observable in the SAED pattern in Fig. 21(c) as an arc (almost incorporated into the diffuse ring at 1.95 Å) and aligned to the arc at 3.5 Å. These observations indicate a smaller degree of ordering or extension of the graphitic arrangements in IBS films as compared to their DC-MS counterparts.

The less structure $CN_x$ films grown by IBS with respect to MS (compare the micrographs of Fig. 21 with those shown in Fig. 1) can be understood by the XANES and Raman studies presented above. $CN_x$ films grown by IBS and DC-MS present in both cases a significant participation of three-fold nitrogen but a slightly higher relative content of cyanogen-like bonding environments is attained in IBS (see Fig. 20). The detrimental role of nitrile sites in FL arrangements is obvious since their terminating character hinders the further extension of graphene sheets, which could explain the lower degree of FL character in FL-$CN_x$ films grown by IBS with respect to dc-MS. The introduction of $N_2$ ion assistance (100 eV) during IBS was shown to induce a significant reduction of the relative content of nitrile bonding environments [50]. Therefore, the less structured films grown by IBS can be partially attributed to the lack of ion assistance during growth.

## 6. FL CARBON FILMS BY ENCAPSULATION OF METAL CATALYST NANOPARTICLES

In previous sections, we have focused our study on the growth of FL-$CN_x$ due to the lower energy barrier for the formation of such arrangements upon N incorporation. In this section we present other route for the yield of FL carbon-based thin films by means of incorporation of metal catalysts nanoparticles during the growth. With the same argument, the presence of the metal catalysts reduces the energy barrier to induce curvature in graphitic planes in a similar way than in the process used for CNT growth and nucleation.

The early investigations of the arc evaporation products based on so-called Krätschmer-Huffman carbon arc process of composite anode [77] showed that, in addition to the $C_{60}$, endohedral fullerenes, nanotubes, carbon onions and carbon soot, there are additional by-product structures which consist of polyhedral C nanoparticles with internal cavities filled with metal or metal carbides [78,79,80,81]. Numerous metallic elements have been reported to be encapsulated in such cage structures [82]. The discovery of this encapsulation mechanism paved a new research pathway in FL materials due to two reasons. First, it provides a possibility to study very small amounts of the encapsulated material which can potentially reveal new properties only available at the nanoscale [82]. For example, the encapsulated nanoparticles of ferromagnetic materials are predominantly monodomain and show the full range of nanoparticle magnetic phenomena such as superparamagnetism and hysteresis below blocking temperature [83,84]. Second, the protective nature of the encapsulating graphite against degradation promised applications such as nanowires, magnetic data storage, ferrofluids, magnetic toner for xerography, or as contrast agents in magnetic resonance imaging [82,85].

However, the main drawback of the above arc process is that it produces a full spectrum of carbon nanostructures and the ratio of the carbon encapsulating nanoparticles is low. In addition, this process provides little control over the particle size, and there is considerable carbonaceous debris which is difficult to remove. Another arc evaporation method was introduced by Dravid *et al.* [86], which permits to control significantly better the nanoparticles size distribution. However, the elimination of carbon by-products still remained a challenge by employing this method.

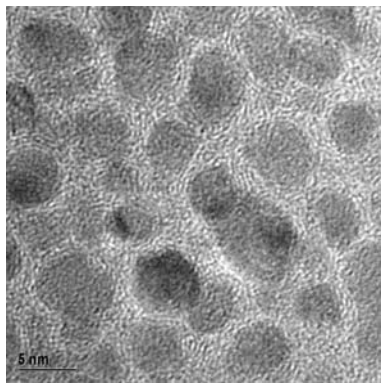

**Figure 22**. HRTEM image showing the encapsulation of Ni nanoparticles in a C film produced by IBS [87]. The C matrix shows the formation of clear FL features (FL-C).

A different approach was introduced by Hayashi *et al.* [88] where cobalt was co-sputtered with carbon by using a conventional ion beam sputtering technique. The subsequent annealing of the as-grown films results in the formation of cobalt nanoparticles wrapped in curved graphitic planes. This approach permits to combine the protective and the decoupling properties of the carbon in a single film. This work inspired numerous studies on the encapsulation of metallic nanoparticles in thin films. Different methods have been later successfully employed to synthesize such films like filtered cathodic vacuum arc [89], plasma enhanced chemical vapour deposition [90], ion beam sputtering [91,92,93,94] or magnetron sputtering [95,96,97,98]. Different structures have been reported starting from metal nanocolumns wrapped in standing graphitic planes to completely enclosed metal nanoparticles of elongated or spherical shapes. As an illustrative example, Figure 22 shows the encapsulation of Ni nanoparticles in a C matrix produced by IBS. It is relevant the formation of clear FL features in the C region.

Besides the magnetic properties, thin films composed of metal nanoparticles has shown fairly good tribological properties, which make them to be promising materials as low-friction, self-lubricating, wear-resistant coatings [97,99]. In addition, the cage structure makes them also possible candidates to be nanocontainers for foreign materials [96,97], while the encapsulation of the metallic particles in the dielectric material makes them a promising candidate as 2D array of tunnel junctions or as nanoscale photonic devices [98,100]. In addition, the encapsulating property has been shown not only for pure C but for other layered materials such as $CN_x$ [97] or hexagonal boron nitride (h-BN) [101,102,103,104]. The latter is of particular interest because in contrast to C or $CN_x$, h-BN is an electrical insulator while it has very similar lattice structure as that of graphite. This property is of particular significance if one considers the electron transport in future nanoscale devices.

Despite the successful synthesis of different encapsulated structures, there are still several major challenges which should be overcome for these materials to be considered for applications. First of all, the control of particle size, since it controls the properties of the overall material. The recent research developments in this field show that narrower and narrower distributions are achieved; however for most applications these distributions in size are still not satisfactory. Second, the control of the interdistance between nanoparticles or the production of ordered arrays remains a major problem. This property is not so important for mechanical applications, while it is crucial for electronical or optical applications. Following this, new physical, chemical or technological approaches are needed to be introduced.

## 7. CONCLUSION

As a summary, we have studied the bonding structure of FL-$CN_x$ thin films by comparison of samples with different degree of FL character. In this way, we have gained a considerable insight in the basic understanding of these carbon-based nanostructured coatings, such as the paths for in-plane curvature and cross-linking mechanisms. Our results have shown that FL-$CN_x$ films are attained by the promotion of three-fold N (substitutional and pyrrole-like) over low coordination (nitrile- and pyridine-like) sites. It has been found that π bonding around N atoms is greatly affected by the formation of curved graphitic structures, suggesting that N sites may act as cross-liking sites or originating curvature.

By comparing $CN_x$ films grown under different methods, we have also obtained relevant information about the growth mechanisms leading to the formation of FL arrangements. In particular, we have seen that low-energy ion bombardment is not the only requisite for the microstructural evolution and that the polyatomic $C_xN_y$ species emitted during the sputtering of C targets can significantly participate in the film growth. Also, the presence of low-energy ion bombardment seems to be necessary for the development of well-structured FL-$CN_x$ films by promotion of favorable sites due to preferential sputtering or by inducing surface diffusion processes. However, excessive energetic conditions induce the amorphization and disruption of the growing microstructure.

Finally, an additional path for the promotion of FL arrangement in pure carbon films is presented. This process is based on the catalytic character of metal nanoparticles that induce the formation of carbon nanostructures such as occurs in CNT growth. In this way, embedding metal catalysts nanoparticles during the growth can lead to the growth of FL films with new and interesting properties coming from the microstructure and nanostructuration process.

## ACKNOWLEDGMENTS

The work reviewed in this chapter has been supported by the EU and carried out inside the *"Synthesis, structure and properties of new fullerene-like materials"* IHP-Network (Ref: HPRN-CT-2002-00209) and *"Fullerene-based Opportunities for Robust Engineering: Making Optimised Surfaces for Tribology (FOREMOST)"* project (Ref: NMP3-CT-2005-515840). We would like to thank J. Neidhardt and L. Hultman for providing DC-MS $CN_x$ samples and fruitful discussions.

## MAIN PUBLICATIONS BY THE AUTHORS RELATED TO THIS CHAPTER

Jiménez, I.; Gago, R.; Albella, J. M.; Caceres, D.; Vergara, I.; *"Spectroscopy of p bonding in hard graphitic carbon nitride films: Superstructure of basal planes and hardening mechanisms"*, *Phys. Rev. B* 2000, 62, 4261-4264 (Ref. 23).

Jiménez, I.; Gago, R.; Albella, J.M.; Terminello, L.J. *"X-Ray absorption studies of bonding environments in graphitic carbon nitride"*, *Diam. Relat. Mater.* 2001, 10, 1166-1169 (Ref. 39).

Gago, R.; Jiménez, I.; Neidhardt, J.; Abendroth, B.; Caretti, I.; Hultman, L.; Möller, W. *"Correlation between bonding structure and microstructure in fullerene-like carbon nitride thin films"*, *Phys. Rev. B* 2005, 71, 1254514 (Ref 21).

Gago, R.; Neidhardt, J.; Vinnichenko, M.; Kreissig, U.; Czigány, Zs.; Kolitsch, A.; Hultman, L.; Möller, W. *"Synthesis of carbon nitride thin films by low-energy ion beam assisted evaporation: on the mechanisms for fullerene-like microstructure formation"*, *Thin Solid Films* 2005, 483, 89-94 (Ref. 45).

Abrasonis, G.; Gago, R.; Jiménez, I.; Kreissig, U.; Kolitsch, A.; Möller, W. *"Nitrogen incorporation in carbon nitride films produced by direct and dual ion beam sputtering"*, *J. Appl. Phys.* 2005, 98, 074907 (Ref. 50).

Gago, R.; Abrasonis, G.; Mücklich, A.; Möller, W.; Czigány Zs.; Radnóczi, G. *"Fullerene-like arrangements in carbon nitride thin films grown by direct ion beam sputtering"*, *Appl. Phys. Lett.* 2005, 87, 071901 (Ref 76).

Abrasonis, G.; Gago, R.; Vinnichenko, M.; Kreissig, U.; Kolitsch, A.; Möller, W. *"Six-fold ring clustering in sp2-dominated carbon and carbon nitride thin films: a Raman spectroscopy study"*, *Phys. Rev. B* 2006, 73, 125427 (Ref. 56).

## LIST OF REFERENCES